\documentclass{article}
\usepackage[T1]{fontenc} 
\usepackage[utf8]{inputenc} 
\usepackage{ismir,amsmath,cite,url}
\usepackage{graphicx}
\usepackage{color}
\usepackage{subfig}
\usepackage{booktabs}
\usepackage{microtype}

\usepackage{xcolor}

\title{Beat this!\\Accurate beat tracking without DBN postprocessing}

\multauthor{Francesco Foscarin$^{*1,2}\thanks{* Equal contribution.}$ \hspace{1.5cm}Jan Schl\"{u}ter$^{*1}$ \hspace{1.5cm} Gerhard Widmer$^{1,2}$}
{$^1$ Johannes Kepler University, Linz, Austria\\
$^2$ LIT AI Lab, Linz Institute of Technology, Austria\\
{\tt firstname.lastname@jku.at}}

\def\authorname{F. Foscarin, J. Schl\"{u}ter, and G. Widmer}

\usepackage[bookmarks=false,pdfauthor={\authorname},pdfsubject={\papersubject},hidelinks]{hyperref}

\begin{document}

\maketitle
\begin{abstract}
We propose a system for tracking beats and downbeats with two objectives: generality across a diverse music range, and high accuracy.
We achieve generality by training on multiple datasets -- including solo instrument recordings, pieces with time signature changes, and classical music with high tempo variations -- and by removing the commonly used Dynamic Bayesian Network (DBN) postprocessing, which introduces constraints on the meter and tempo.
For high accuracy, among other improvements, we develop a loss function tolerant to small time shifts of annotations, and an architecture alternating convolutions with transformers either over frequency or time.
Our system surpasses the current state of the art in F1 score despite using
no DBN.
However, it can still fail,
especially for difficult and underrepresented genres, and performs 
worse on continuity metrics, so we publish our model, code, and preprocessed datasets, and invite others to beat this.

\end{abstract}
\section{Introduction}\label{sec:introduction}

Beat tracking is the task of estimating the temporal locations of musical beats in an audio signal. It is often combined with the downbeat tracking task, which targets a higher metrical level: tracking the beginning of each measure.
Despite being one of the long-standing problems in the Music Information Retrieval (MIR) field, it still attracts attention and several approaches were proposed in recent years~\cite{lowreshighres2023,nodbn2022,beattransformer2022,beatstructurekim2023,chiu2023local,Maia2023AdaptingMT,cmmr2023,desblancs2023zero}. Most recent work follows a common pipeline: the audio files are transformed into some spectrogram-like representation, then a deep neural network predicts frame-wise beat and downbeat probabilities, which are postprocessed to obtain the final predictions. The most widely used postprocessing technique is the Dynamic Bayesian Network (DBN) in the form proposed by B{\"o}ck et al.~\cite{bock2016joint}. It addresses four tasks: variable threshold peak-picking, forcing the tempo to stay in a certain range (i.e., limiting the allowed distance between beats), limiting sudden tempo changes, and (for downbeat tracking) ensuring that the downbeat falls every $n$ beats, where $n$ is constant for a piece of music and is selected from a limited list of values.

We argue that the DBN is a problematic component because it is inherently bound to fail for several music pieces: pieces with time signature changes, pieces whose tempo falls outside of the tempo range (or that slow down/speed up outside the tempo range), and pieces whose number of beats per measure are not included in the list of supported values. 
Moreover, it has a fixed parameter controlling allowed tempo variations, although we can expect, for example, classical music to have bigger tempo variability than rock music. 
Finally, even the hypotheses of having periodic beats and downbeats may be invalid, for example, for songs where the players make mistakes or audio tracks containing multiple concatenated songs.

Still, the DBN performs well on most pieces commonly used to train and evaluate beat tracking systems: music with a constant time signature of 3/4 or 4/4 and a stable, medium tempo. This can be seen from the default DBN parameters which most systems use,\footnote{We could verify that \cite{beattransformer2022,beatstructurekim2023,desblancs2023zero} use these parameters, since their code is publicly available, and we assume \cite{bytedancesoabt2022,lowreshighres2023} do as well, since they do not mention any details in their paper.} i.e., tempo range $[55, 215]$ BPM, beats per measure $[3,4]$, and a tempo variability optimised on pop, rock and dance datasets. Pieces outside these specifications are likely to be mispredicted by the system, but form a minority in typical datasets. Therefore, in terms of evaluation metrics, it usually does not pay to remove the DBN.
However, working in these ``simplified'' conditions blocks research from solving corner cases in existing datasets and targeting more challenging or diverse data.

Our first goal is thus to replace the DBN with minimal postprocessing, free of the aforementioned musical assumptions.
A recent attempt to remove the DBN was made by Chen and Su~\cite{nodbn2022}. However, their system may not look appealing to practitioners requiring beat tracking for downstream tasks, or researchers seeking a system to improve, as its accuracy falls clearly behind DBN-based ones.

Our second goal is to provide a powerful basis for practitioners and researchers.
The current best-performing system (which uses a DBN) from Hung et al.~\cite{bytedancesoabt2022} falls short in this regard, as its code or a pretrained model is not public, its architecture is very complex, and (to the best of our knowledge) the results could not be reproduced by others.

In this paper, we present an open-source system that obtains new state-of-the-art F1 scores without a DBN.
It is based on a rotary transformer \cite{su2024roformer} applied on spectrograms, with the following novelties:
(1) We design a frontend alternating convolutions with a transformer variant by Lu et al.~\cite{lu2024music} that attends alternatively over frequency bins or time frames.
(2) We train with a shift-tolerant binary cross-entropy (BCE) that can cope with small deviations in the beat/downbeat annotations, and with weights on beat/downbeat frames to balance their relative scarcity.
(3) We propose an approach that encourages downbeat predictions to be a subset of beat predictions, and
(4) a data augmentation masking input segments to encourage the network to consider a longer context.
 Our code, pretrained models, and preprocessed datasets are openly available.\footnote{\url{https://github.com/CPJKU/beat_this}}

\section{Related work}\label{sec:related}
The currently best-performing model (on the GTZAN dataset~\cite{gtzan} commonly used for evaluation) is by Hung et al.\cite{bytedancesoabt2022} and serves as a point of comparison. It uses a complex neural network architecture named SpecTNT which alternates computing frequency-related features with a frequency-oriented transformer, and processing a virtual extra frequency band with a time-oriented transformer. This runs in parallel with a more widely used Temporal Convolutional Network (TCN, a fully convolutional network with dilated convolutions), and the outputs of the two networks are merged for the final predictions. Unfortunately, the approach is not open source, and to the best of our knowledge, no other research group has managed to reproduce its results. Moreover, it still uses the DBN, which, as argued in the introduction, limits the system's generality.

Although no other work could reach the accuracy reported by Hung et at., two other recent beat tracking papers brought new interesting ideas~\cite{beattransformer2022,beatstructurekim2023}. Both perform instrument separation (with a pretrained network) and feed the separate stems 
(bass, drums, vocals, other, for~\cite{beattransformer2022} also piano)
into the model, mixing their information in cross-instrument attention blocks. While this approach is very reasonable from a music perception standpoint, it reduces the generality of the system, since it assumes that the input pieces will contain such instruments, at least to some extent. Another proposal of \cite{beattransformer2022,beatstructurekim2023} is the use of dilated attention, following the successful use of dilated convolutions in beat-tracking architectures to 
increase the receptive field without adding computations. We find that flash attention~\cite{dao2022flashattention} enables us to train with dense attention over a satisfactorily large input size, and leave experiments with dilated attention for future work.

Chen and Su~\cite{nodbn2022} try to remove the DBN and propose a set of improvements. The most impactful is to replace the BCE with the Dice~\cite{milletari2016v} and Focal~\cite{lin2017focal} loss, inspired by their common usage for medical image segmentation. While these losses improve results,
possibly due to their inherent ability to handle unbalanced classes,
we found that a BCE with weights on the positive (beat and downbeat) classes outperforms them. We suspect this is because, in contrast to medical image segmentation, our positive examples are single frames, and the Dice and Focal loss perform better when the area of positive predictions is larger~\cite{abraham2019novel}. 
Another proposal by Chen and Su is to predict the phase of the beat/downbeat instead of a single binary value, following~\cite{oyama2021phase}. Although this seems promising, the results on both papers (and ours) do not show any consistent improvement.

Many recent approaches~\cite{bock2020deconstruct,oyama2021phase,nodbn2022,beattransformer2022} use the additional task of tempo prediction (a single tempo target for each input excerpt) in a multi-task setting. While this improves their results, it goes against our goals of generality, since it assumes an (almost) constant tempo in the file excerpt, which is not the case for many kinds of music.

Other recent papers do not align with the goal of this paper: \cite{lowreshighres2023} explores the usage of different time resolutions between the input audio and output predictions (only addressing beats); \cite{desblancs2023zero} performs unsupervised beat tracking; \cite{Chang2023BEASTOJ} focuses on online beat tracking; ~\cite{chiu2023local} notices the problems of the DBN for music with tempo changes, and proposes a different postprocessing method targeted specifically to classical music; \cite{cmmr2023} focuses on fine-tuning existing systems, and changing the DBN parameters for targeting specific underrepresented genres.


\section{Method}\label{sec:method}
Our beat tracker
is based on a neural network with $\sim$20\,M parameters.
It
starts from 30 seconds of mono audio sampled at 22.05\,kHz and converts it to a 128-bin mel spectrogram from 30\,Hz to 10\,kHz, with a window size of 1024 and hop size of 441 samples (yielding 50 frames per second), and magnitudes scaled via $\ln(1 + 1000 x)$ (similar to $\ln(\max(10^{-3}, x))$, but maps silence to zero).
These hyperparameters were optimised in preliminary experiments.
Our model processes this into frame-wise beat and downbeat probabilities, followed by minimal post-processing to derive beat and downbeat locations.

\subsection{Model}
\begin{figure}
    \centering
    \includegraphics[width=0.90\columnwidth]{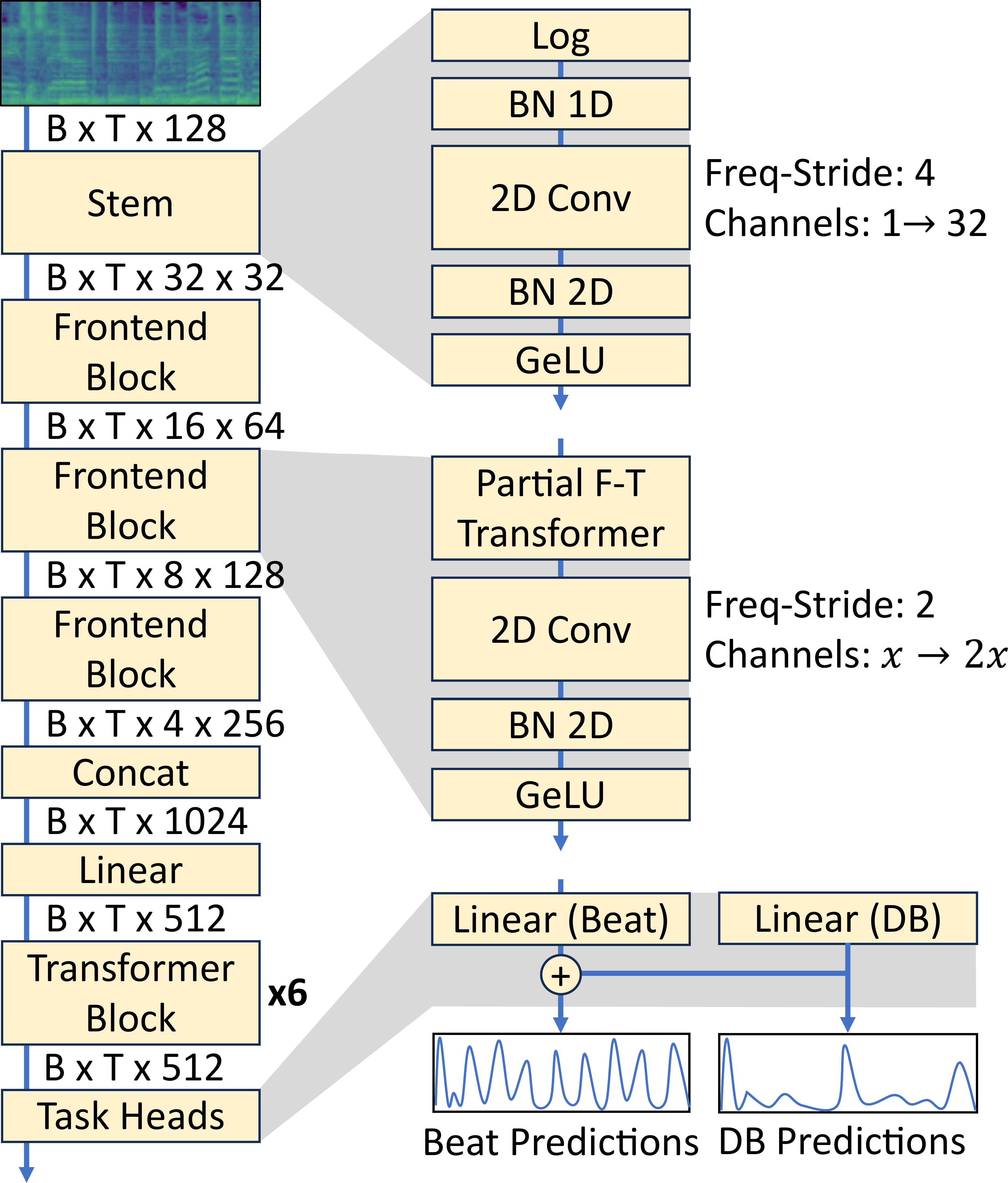}
    \caption{Full model architecture. 
    }
    \label{fig:model}
\end{figure}

Our model (Figure~\ref{fig:model}) processes a $T\!\times\!128$ spectrogram into $T\!\times\!2$ probabilities; $T$ being the number of input frames (1500 in case of 30\,s).
It consists of three components: a frontend converting the spectrogram into a sequence of feature vectors, a transformer processing these vectors, and two task heads computing the output probabilities.

\subsubsection{Frontend}
The frontend's role is to integrate information across the 128 frequency bands into feature vectors.
Typically, this is done via 2d convolutions gradually reducing the number of bands to 1 while increasing the number of channels \cite{bytedancesoabt2022,beattransformer2022}.
We adopt this, but found it helps to interleave convolutions with \textit{Partial Transformers}, which treat the time and frequency axis independently.
Overall, our frontend consists of a stem, three identical blocks, and a linear projection.

The stem (Figure~\ref{fig:model}, top right) starts with a batch normalisation that processes each frequency band separately to homogenise them, followed by a 2d convolution of $3\!\times\!4$ kernels, regular batch normalisation, and GeLU nonlinearity. The convolution is strided to reduce the number of frequency bands to a fourth and creates 32 channels.

Each block (Figure~\ref{fig:model}, middle right) consists of two partial transformers, a strided 2d convolution halving the number of bands while doubling the number of channels, batch normalisation and GeLU.
The first partial transformer is \textit{frequency-directed}, i.e., it processes the $T\!\times\!F\!\times\!C$ tensor by treating each time frame as a sequence of length $F$ (the number of bands), the second one is \textit{time directed} and treats each frequency band as a sequence of length $T$ (the number of frames), an idea adopted from the Band Split RoFormer \cite{lu2024music}.
Each transformer has a head size of 32 (one head in the first frontend block, two in the second, four in the third), rotary positional embedding \cite{su2024roformer}, a sigmoid gate per head \cite[Sec.~4.2]{bondarenko2023quantizable},
and includes a usual pointwise feedforward network with a hidden size of four times the channel count.


After three frontend blocks, the resulting $T\!\times\!4\!\times\!256$ tensor (4 bands, 256 channels) is reshaped to a $T\!\times\!1024$ tensor and linearly projected to $512$ features.

\subsubsection{Transformer}
The transformer makes up the bulk of our model's parameters and compute.
It consists of 6 stacked transformer blocks processing the 512-dimensional vectors with 16 heads of size 32, rotary positional embedding, sigmoid gating, and a pointwise feedforward network of 2048 hidden units.
This matches the configuration in the frontend transformer blocks, but as it processes a single sequence of 512-dimensional feature vectors, it is a regular transformer over time without separately considering a frequency dimension.
Its goal is to map the 512 input features to a space that relates to beats and downbeats.
Due to the attention mechanism, the transformer's receptive field covers the full sequence, and it could therefore produce an output that has characteristics that we want in the beat predictions, for example, regularity.

\subsubsection{Task Heads}\label{sec:task_heads}
The output of the final transformer block is processed by two linear layers, one for beats and one for downbeats. Initially, we used the common approach of passing their output into 2 sigmoid functions to produce a probability for each input frame, and then threshold this probability at $0.5$ to obtain "hard" beat predictions. However, we observe that this sometimes produces downbeat predictions not coinciding with a beat prediction, which is allowed under the evaluation metrics but is a musically invalid and unusable output.
This problem is solved when using a DBN to jointly process beats and downbeats. However, we noticed that several works, e.g.,~\cite{beattransformer2022,Maia2023AdaptingMT}, use two independent DBNs to predict beats and downbeats (and others~\cite{nodbn2022,bytedancesoabt2022} do not specify). To our surprise, this leads to better metrics, but it severely limits practical use.

To mitigate this problem, we propose a \textit{Sum Head} that sums the output of the beat and downbeat layers, and treats this as the beat logits (for prediction and training). This is a very simple way of helping the network produce a beat when there is a downbeat (though it does not enforce that; a highly negative output of the beat layer can still counter the downbeat layer). We explored other ways of aggregating the beat and downbeat logits, like taking their maximum, but this hampered training due to the sparser gradients. On the GTZAN dataset, the sum head almost halves the percentage of downbeats that are more than 70\,ms away from the closest beat, from 1.1\% to 0.62\%, compared to directly using the output of the linear layers. We observe that the remaining unmatched downbeats are in pieces with very erratic predictions that would be incorrect anyway.

Some systems circumvent the problem by using a 3-way classifier (beat vs.\ downbeat vs.\ none) instead of the two binary classifiers (beat vs.\ none, and downbeat vs.\ none). However, to be able to train on datasets that do not include downbeat annotations, we stick to binary classifiers.


\subsection{Postprocessing}
To obtain beat/downbeat locations, we pick all frames assigned the highest beat/downbeat probability within a neighborhood of $\pm3$ frames ($\pm 70$\,ms), and probability > 0.5. In case adjacent frames are assigned the same probability, we report their center. Finally, we move all downbeat predictions to the closest beat prediction to correct the remaining mismatches described in the previous section.
For music pieces longer than 30\,s, we concatenate predictions over non-overlapping 30-second excerpts.

\subsection{Loss}\label{sec:loss}

\begin{figure}
    \centering
    \includegraphics[width=\columnwidth]{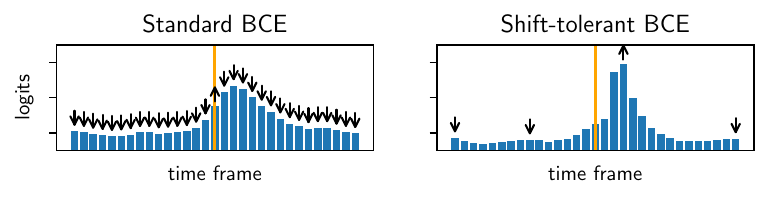}
    \caption{The standard binary cross-entropy loss (left plot) encourages high network outputs (upward arrow) at beat annotations (vertical line), and low outputs for all other frames (downward arrows). Max-pooling the predictions over time redistributes gradients to local maxima (right plot). This way, slightly shifted annotations do not affect learning, and the network produces confident sharp peaks.}
    \label{fig:losses}
\end{figure}

The model is trained by gradient descent on a loss function that compares the frame-wise beat and downbeat predictions with frame-wise binary annotations.
The usual loss for binary classification is Binary Cross-Entropy, $L_{\text{bce}}(\mathbf{y}, \mathbf{\hat{y}}) = -\sum_t y_t \log(\hat{y}_t) + (1-y_t) \log(1-\hat{y}_t)$.
Training with BCE leads to unconfident predictions since the problem is heavily imbalanced.
To counter this, we can weight positive examples by a factor $w$ as $L_{\text{wbce}}(\mathbf{y}, \mathbf{\hat{y}}, w) = -\sum_t w y_t \log(\hat{y}_t) + (1-y_t) \log(1-\hat{y}_t)$.
We found that setting $w$ to the number of negative examples divided by the number of positive examples (over the training set) yields the best result and is crucial when not using a DBN.

Another problem persists: annotations are not precise down to our spectrogram resolution, due to annotators' impreciseness, players' asyncronicity, or simply the limits of human perception.
This is taken into account during the evaluation, e.g., the typically used F1 score accepts predictions in a $\pm 70$ ms window around labels.
During training, the BCE loss punishes close positive predictions (Figure~\ref{fig:losses}, left) even though they may be correct, thus creating two problems: training is slower and the network learns to predict wide ``blurred'' peaks.
This is commonly addressed by adding two extra positive labels around each annotation weighted by 0.5, but this only mitigates the former problem without helping with the latter.
Instead, we max-pool predictions over time (7 frames, stride 1) before comparing them to the labels.
In this way, only the largest positive prediction $\pm 3$ frames from each label is considered (Figure~\ref{fig:losses}, right).
The loss for negative examples is ignored $\pm 6$ frames from each label, as this is how far a max-pooled prediction $3$ frames away from a label spreads.
Denoting max-pooling of $k$ frames with $m_k(\cdot)$, we can formalise our \textit{Shift-tolerant weighted BCE} as:
$L_{st}(\mathbf{y}, \mathbf{\hat{y}}, w) = -\sum_t w y_t \log(m_7(\mathbf{\hat{y}})_t) + (1 - m_{13}(\mathbf{y})_t) \log(1 - m_7(\mathbf{\hat{y}})_t)$.

\subsection{Data Augmentation}\label{sec:augmentation}

\textbf{Masking.} To encourage the model to not only rely on local information for its predictions, we mask 0 to 6 areas of 0.5 to 2\,s. Each masked area is randomly divided into 5 to 10 parts which are randomly reordered. This destroys local correspondence between audio and beats without changing local input statistics, and works better than zero masking as employed in SpecAugment~\cite{Park2019SpecAugmentAS}. We assume our approach makes it harder for the network to learn a dedicated behaviour for masked areas.

\noindent
\textbf{Pitch and time.} We speed up and slow down every song by 20, 16, 12, 8, and 4\%, and transpose by at most +6 and $-$5 semitones.
We precompute these augmentations (using Pedalboard~\cite{sobot_peter_2023_7817838}) so experiments become reproducible without access to the original audio.
To limit storage use, tempo and pitch augmentations are not combined, giving 22 variations for each song.
We verified that our limited tempo augmentation gives comparable results to the commonly used approach by Böck and Davies~\cite{bock2020deconstruct} of performing on-the-fly augmentations by randomly changing the hop size of the STFT, at the advantage of not requiring audio access.


\section{Experiments}\label{subsec:experiments}
We perform 8-fold cross-validation experiments on multiple datasets, compute results on the test-only GTZAN dataset, and do an ablation study.

We use the standard beat-tracking metrics: F1, CMLt, and AMLt 
and compute them using the \texttt{mir\_eval} package~\cite{raffel2014mir_eval} with default parameters.\footnote{This includes ``trim\_beats'' of 5s that discards the first 5 seconds during the evaluation, which is a choice we do not necessarily approve of, but we use it to be consistent with what seems the standard way of evaluating.} CMLt and AMLt are called continuity metrics, and only consider a beat/downbeat as correct if both it and the previous beat/downbeat are correct; AMLt also accepts different metrical levels such as half or double time, and offbeats~\cite{tempobeatdownbeat:tutorial}.
The metrics and their settings match those used by Hung et al.~\cite{bytedancesoabt2022}; this enables a comparison with their reported results, though it is unclear which 8-fold datasplit they use,
and we cannot run any statistical significance tests since their code is not reproducible. Therefore, any comparison needs to be taken only as an indication.

\subsection{Datasets}\label{sec:dataset}
We train and validate with several datasets: Simac\cite{simac_ballroom}, SMC~\cite{smc},
Hainsworth~\cite{hainsworth2004}, Ballroom~\cite{simac_ballroom,ballroom}, HJDB~\cite{hockman2012one}, Beatles~\cite{beatles}, Harmonix~\cite{harmonix}, 
RWC~\cite{rwc,rwc2} (classical, pop, royalty-free, and jazz), 
TapCorrect~\cite{driedger2019towards}, JAAH~\cite{jaah}, Filosax~\cite{filosax}, ASAP~\cite{foscarin2020asap}, Groove MIDI~\cite{groove_midi}, GuitarSet~\cite{guitarset}, Candombe~\cite{candombe}. The first two datasets contain only beat annotations, all others both beat and downbeats. We discard one Beatles piece which does not contain downbeat annotations and one with empty beat annotations, resulting in a total of 4556 tracks. For comparison, Hung et al.~\cite{bytedancesoabt2022} train with only the first 7 datasets reported above (Simac to Harmonix), plus RWC pop, for a total of 3144 pieces (when assuming the same handling of missing annotations).
We use the GTZAN~\cite{gtzan} dataset (993 pieces discarding one unannotated track and 6 tracks that miss downbeat annotations) for testing only.

We use only the backing tracks of Filosax without the saxophone solos. For ASAP, we discard the tracks that contain the ``rubato'' beat annotations. In Groove MIDI, we keep all pieces that are longer than 20 seconds and use the provided audio renderings. We only use the comping tracks of GuitarSet, discarding the solos. 

We employ the \textit{8-fold cross-validation splits} published by Böck and Davies~\cite{bock2020deconstruct} for the datasets they used and produce new ones for
our added datasets, ensuring different versions of the same piece are not spread across folds, and stratifying by metadata when possible.
We also produce a new \textit{single split} with $\sim 15\%$ of the pieces on each dataset as validation (again taking care of different versions of the same piece).

\subsection{Training}
We train for 100 epochs with gradient accumulation over 8 batches of size 8,\footnote{This enables training with under 8\,GiB of GPU memory.} AdamW optimizer~\cite{adamw}, weight decay of 0.01 (excluding biases and learned norms), learning rate warm-up~\cite{huang2020improving} of 1000 steps to a maximum of 0.0008, and cosine annealing.
During each epoch, we randomly sample 30 seconds of each piece, and pad if the total piece duration $l$ is less than 30 seconds. 
We draw $k$ samples from pieces that are longer than 30 seconds, following the equation $k = \textit{round} \left(\alpha \cdot l/30 \right)$
with $\alpha=0.65$, since we observe it leads to faster convergence than using one random sample per piece, or $l/30$ non-overlapping samples. On average, this yields $\sim 3$ samples per piece. During training, every time a sample is drawn, we randomly select a precomputed augmentation described in Section~\ref{sec:augmentation}, and apply masking.
The full training takes around 8 hours on a single NVIDIA RTX~2080~Ti, 6 hours on A40, and 4 hours on A100.

During our experimentation, we found that to achieve good results without a DBN, we need our network to be overconfident in its predictions.
This may seem to violate usual deep learning practice, but can be explained by a closer look at the beat tracker's desirable output. We do not want our network to produce probabilities close to 0.5 when unsure, since this will lead to random oscillations between positive and negative predictions, and thus erratic beats. Instead, we want it to give steady, high-probability predictions even when unsure, exactly like the DBN would.
To achieve this, we keep training even after the validation loss starts increasing,
which would typically indicate overfitting.
Indeed, we see that the validation F1 score continues to improve even with increasing validation loss.
This means that even with our modifications, the BCE loss is not a good indicator of the F1 score, and further research into alternative losses may be valuable.

The reader may wonder why, once we obtain our well-performing network, we do not use the DBN to increase the metrics even more. 
By having overconfident predictions, we reduce the benefits of such a postprocessing method. With a degree of simplification, we can imagine the DBN as using a model's most high-confident predictions to infer beats in low-confidence areas. By avoiding the low-confidence predictions, we are disrupting this mechanism. 

\subsection{Main Results}

We report the results on our 18 datasets in the commonly used 8-fold cross-validation setting: each dataset is split into 8 parts, we jointly train on $\text{18}\cdot\text{7}$ parts (all but one per dataset) and predict on the remaining 18, after 8 such runs we covered all pieces and average metrics over pieces by dataset. We observe that our model outperforms Hung et al.~\cite{bytedancesoabt2022}, except for Harmonix and RWC Pop (downbeat). 
In our results, the lowest downbeat performance is obtained in the ASAP and RWC Classical datasets, confirming the well-known difficulty of beat-tracking classical music~\cite{nodbn2022,chiu2023local}. Performance on SMC (where only beat annotations are accessible) is also very low, consistent with the outcomes of other systems, highlighting the substantial room for improvement that beat tracking systems continue to hold.

\begin{table}
\centering
\begin{tabular}{lcccc}
\toprule
 & \multicolumn{2}{c}{Beat F1} & \multicolumn{2}{c}{Downbeat F1}\\
\cmidrule(lr){2-3}
         \cmidrule(lr){4-5} 
 & Our & Hung  & Our & Hung \\
 \midrule
ASAP & 76.3 & - & 61.2 & - \\
Ballroom & \textbf{97.5} & 96.2 & \textbf{95.3} & 93.7 \\
Beatles & \textbf{94.5} & 94.3 & \textbf{88.8} & 87.0 \\
Candombe & 99.7 & - & 99.7 & - \\
Filosax & 99.5 & - & 98.5 & - \\
Groove MIDI & 93.7 & - & 82.1 & - \\
GuitarSet & 92.0 & - & 88.1 & - \\
Hainsworth & \textbf{91.9} & 87.7 & \textbf{80.0} & 74.8 \\
Harmonix & \textbf{95.8} & 95.3 & 90.7 & \textbf{90.8} \\
HJDB & 98.2 & - & 96.6 & - \\
JAAH & 95.1 & - & 85.0 & - \\
RWC Classical & 77.1 & - & 66.3 & - \\
RWC Jazz & 83.3 & - & 80.7 & - \\
RWC Pop & \textbf{96.1} & 95.0 & 93.7 & \textbf{94.5} \\
RWC RF & 94.5 & - & 91.9 & - \\
Simac & 77.9 & - & - & - \\
SMC & \textbf{62.7} & 60.5 & - & - \\
TapCorrect & 93.0 & - & 86.4 & - \\
\bottomrule
\end{tabular}
\label{tab:cv}
\caption{Results with 8-fold cross-validation.}
\end{table}

\begin{table*}[h!]
    \centering
    \begin{tabular}{lllllll}
        \toprule
         & \multicolumn{3}{c}{Beat} & \multicolumn{3}{c}{Downbeat}\\
         \cmidrule(lr){2-4}
         \cmidrule(lr){5-7} 
         & \multicolumn{1}{c}{F1} & \multicolumn{1}{c}{CMLt} & \multicolumn{1}{c}{AMLt} & \multicolumn{1}{c}{F1} & \multicolumn{1}{c}{CMLt} & \multicolumn{1}{c}{AMLt} \\
         \midrule
         Hung et al.~\cite{bytedancesoabt2022} & $88.7$  & $\textbf{81.2}$ & $ \textbf{92.0}$ & $75.6$ & $71.5$ & $\textbf{88.1}$\\
         \textbf{Our system} & $\textbf{89.1} \pm \textbf{0.3} $ &  $79.8 \pm 0.6 $ &  $89.8 \pm 0.4 $ &  $\textbf{78.3} \pm \textbf{0.4} $ &  $67.3 \pm 0.8 $ &  $79.1 \pm 0.6 $  \\
         -- limited to data of \cite{bytedancesoabt2022} & $88.9 \pm 0.1 $ &  $79.9 \pm 0.4 $ &  $89.4 \pm 0.2 $ &  $75.5 \pm 0.5 $ &  $60.8 \pm 1.2 $ &  $75.5 \pm 0.5 $  \\
         -- smaller model & $88.8 \pm 0.2 $ &  $79.4 \pm 0.4 $ &  $89.0 \pm 0.4 $ &  $77.2 \pm 0.2 $ &  $65.3 \pm 0.3 $ &  $78.0 \pm 0.3 $  \\
         -- with DBN & $88.1 \pm 0.3 $ &  $80.5 \pm 0.4 $ &  $91.1 \pm 0.2 $ &  $77.4 \pm 0.2 $ &  $\textbf{73.3} \pm \textbf{0.2} $ &  $87.8 \pm 0.5 $  \\
         \bottomrule
    \end{tabular}
    \caption{Evaluation on the test dataset (GTZAN). The results for Hung~et al.\cite{bytedancesoabt2022} are taken from their paper.}
    \label{tab:gtzan}
\end{table*}

We also report the results on the GTZAN dataset in Table~\ref{tab:gtzan}. We compute these results with a single model trained on the entirety of our training-val dataset (note that we do not perform any early stopping or other techniques for which the validation dataset may still be necessary). All our runs are computed 3 times with different random seeds, and we report the means and standard deviations of the computed metrics over the 3 seeds.
We notice that even when training on the reduced collection of datasets by Hung et al.\ (third row in the table), we still outperform their F1 score without a DBN, proving the effectiveness of our design choices. 
Our main model has $\sim$20\,M parameters, 5 times more than Hung et al.\ with 4\,M, so we also show the results for a smaller model with the hidden dimension of the main transformer blocks reduced from 512 to 128, and the number of heads from 16 to 4. This small model has $\sim$2\,M parameters and still gives SOTA F1 scores.

Disappointingly, we notice that the continuity metrics (CMLt and AMLt) are lower than those of Hung et al. From qualitative inspections of the results, we notice that for complex or underrepresented pieces, our network introduces non-periodic beats, which drastically lower the continuity metrics. We are then brought to wonder why our network cannot learn a supposedly obvious behaviour, such as only producing periodic-like output, and we can propose two explanations. Firstly, our loss does not specifically penalise non-periodic predictions, but treats each beat individually. This results in a discrepancy between what is preferred by continuity metrics and what the network learns to predict in difficult parts to minimise the loss. Secondly, our datasets contain several non-periodic annotations, some due to quality issues (see Section~\ref{sec:quality}), some in correctly annotated pieces such as tapcorrect\_10 or beatles\_Wild-Honey-Pie, where a 2/4 measure in the middle of a 4/4 piece disrupts the assumption of periodicity for downbeats. Finally, one could also question the generality of the AMLt metric as a tool to quantify double/half-time errors, since the computations of different metrical levels assume that the time signature and the tempo do not change and that a measure can always be divided into 2 or 3 parts.

Using a DBN increases our CMLt downbeat performance by correcting some of the (wrongly) non-periodic outputs, but it reduces our F1 performance, by changing other otherwise correct predictions that fall outside the DBN assumptions. The AMLt score does not increase since our network is overconfident in its predictions, and the DBN cannot easily switch to another metrical level.

\subsection{Ablation Studies}

\begin{table}
\centering
\begin{tabular}{lcc}
\toprule
& Beat F1 & Downbeat F1\\
\midrule
\textbf{Our system} & $\textbf{92.6} \pm \textbf{0.1}$ & $\textbf{85.4} \pm \textbf{0.1}$ \\
No sum head & $\textbf{92.6} \pm \textbf{0.1}$ & $85.0 \pm 0.1$ \\
No tempo augmentation & $92.5 \pm 0.1$ & $84.9 \pm 0.1$ \\
No mask augmentation & $92.2 \pm 0.0$ & $84.5 \pm 0.3$ \\
No partial transformers & $92.2 \pm 0.1$ & $83.9 \pm 0.2$ \\
No shift tolerance & $91.2 \pm 0.2$ & $82.2 \pm 0.4$ \\
No pitch augmentation & $88.3 \pm 0.4$ & $80.8 \pm 0.5$ \\
No shift tol., no weights & $79.5 \pm 0.7$ & $68.7 \pm 0.8$ \\
\bottomrule
\end{tabular}
\caption{Ablation studies on the single split validation dataset, ordered by decreasing downbeat F1.}
\label{tab:ablation}
\end{table}

We ablate multiple components of our model on the single split described in Section~\ref{sec:dataset}.
We perform every experiment 3 times with different seeds and report the mean and standard deviation on the validation set in Table~\ref{tab:ablation}.
The usage of our Sum Head shows little impact, but we use it to have a musically valid output, rather than to increase the F1 score. Pitch, mask, and tempo augmentations help, in this order of importance.
The usage of partial transformers in our frontend proves more effective than only having convolutions.
Our most impactful design choice is the weighted shift-tolerant loss. Using a normal BCE with positive example weights results in decreased performance, which decreases even further when the weights are removed.

\subsection{Notes on Dataset Quality}\label{sec:quality}
While exploring the datasets, we found multiple problems in the annotations, and we think this hinders the development of better models, especially for downbeat predictions. Even the GTZAN dataset, which is commonly used for evaluation, is not immune to quality problems. Some of them are evident and not debatable, like jazz\_00000, jazz\_00002, jazz\_00083, blues\_00015, reggae\_00095, classical\_00077, rock\_00067.
Furthermore, there are pieces where even for experts it would be hard to agree on a unique beat and downbeat annotation, like metal\_00081 or classical\_00056, and multiple annotations would be necessary. 
Finally, some pieces question the primary assumption of beat tracking, i.e., that there is a beat/downbeat to track, like pop\_00064, and jazz\_00003.

\section{Conclusions and Outlook}\label{sec:conclusions}
In this paper, we presented a new beat tracking system which obtained a state-of-the-art F1 score on a very diverse set of music, with minimal assumptions about the tempo, time signature, and their changes over time. Remarkably, we do not use the DBN postprocessing, which was employed by all recent high-accuracy models. However, removing the DBN hurts the CMLt and AMLt metrics. A study on how this trade-off affects human perception, alternative metrics, and a direct comparison with DBN-based models on complex pieces is left for future work.

We emphasise that beat tracking is not a solved problem, even for commonly targeted genres such as rock or electronic music, especially for the downbeat tracking task. We provide an open-sourced model that can be used as a starting point, and we invite future researchers to improve it. Potential directions are: reducing the model parameters, developing new losses that enforce periodicity during training, using other data augmentation techniques to make the system more robust to multiple sound conditions, fine-tuning it on specific genres, and training on larger datasets. The contribution of people with musical expertise will also be essential, as we think that correcting the existing commonly used datasets, and producing new annotated data for underrepresented genres is a crucial step for further development.

\section{Acknowledgements}
This work was supported by the European Research Council (ERC) under the EU's Horizon 2020 research \& innovation programme, grant agreement No.\ 101019375 (\textit{Whither Music?}), and the Federal State of Upper Austria (LIT AI Lab). The computational results presented were achieved in part using the Vienna Scientific Cluster (VSC).

\bibliography{literature}

\end{document}